\pdfoutput=1
\documentclass{osa-article}

\journal{oe}

\usepackage{graphicx}
\usepackage{epstopdf}

\begin{document}

\title{High-purity microwave generation using a dual-frequency hybrid integrated semiconductor-dielectric waveguide laser}

\author{Jesse Mak,\authormark{1,*} Albert van Rees,\authormark{1} Rob E. M. Lammerink,\authormark{1} Dimitri Geskus,\authormark{2} Youwen Fan,\authormark{1} Peter J. M. van der Slot,\authormark{1} Chris G. H. Roeloffzen,\authormark{2} and Klaus-J. Boller\authormark{1,3}}

\address{\authormark{1}Laser Physics and Nonlinear Optics Group, MESA+ Institute for Nanotechnology, Department of Science and Technology, University of Twente, Enschede, The Netherlands\\
\authormark{2}LioniX International BV, Enschede, The Netherlands\\
\authormark{3}Optical Technologies Group, Institute of Applied Physics, University of Münster, Münster, Germany}

\email{\authormark{*}j.mak@utwente.nl} 


\begin{abstract*}
We present an integrated semiconductor-dielectric hybrid dual-frequency laser operating in the 1.5 $\mu$m wavelength range for microwave and terahertz (THz) generation. Generating a microwave beat frequency near 11 GHz, we observe a record-narrow intrinsic linewidth as low as about 2 kHz. This is realized by hybrid integration of a single diode amplifier based on indium phosphide (InP) with a long, low-loss silicon nitride (Si$_3$N$_4$) feedback circuit to extend the cavity photon lifetime, resulting in a cavity optical roundtrip length of about 30 cm on a chip. Simultaneous lasing at two frequencies is enabled by introducing an external control parameter for balancing the feedback from two tunable, frequency-selective Vernier mirrors on the Si$_3$N$_4$ chip. Each frequency can be tuned with a wavelength coverage of about 80 nm, potentially allowing for the generation of a broad range of frequencies in the microwave range up to the THz range.
\end{abstract*}

\section{Introduction}
Photonic integrated microwave sources exhibiting low phase noise are of strong interest for applications such as microwave photonics \cite{MarpaungNatPhotonics2019}, time and frequency metrology \cite{MilloApplPhysLett2009}, and coherent radar sensing \cite{GhelfiNature2014}. There are various promising methods to photonically generate microwaves, e.g., via mode-locked lasers \cite{LoOptLett2018}, Kerr frequency combs \cite{LiangNatCommun2015} and opto-electronic oscillators (OEOs) \cite{MalekiNatPhotonics2011}. However, a disadvantage with mode-locking and Kerr combs is the limited tunability of the generated frequency, which is fixed by the cavity roundtrip length. A disadvantage of OEOs is that the generated frequency is limited by the speed of the modulator and photodetector \cite{YaoElectronLett1994}, which sets the maximum frequency to around 100 GHz. An attractive alternative, which we exploit here, is to realize a narrow-linewidth dual-frequency laser to generate a beat note at the difference frequency. This approach enables the generation of a wide range of frequencies in the microwave range up to the terahertz (THz) range via tuning of the two laser lines.

For many applications, it is important to (1) reduce fast frequency noise, i.e., to reduce the intrinsic linewidth (also termed fundamental linewidth, natural linewidth, or Schawlow-Townes linewidth \cite{SchawlowTownesPhysRev1958}), and to (2) reduce slow (technical) noise of the microwave frequency. The latter is typically achieved via active stabilization using optical phase locked loops (OPLLs) \cite{BalakierIEEEJSelTopQuantumElectron2018}, however, reducing the intrinsic linewidth remains an advantage also for active stabilization. Therefore, here our focus is to realize a dual-frequency laser with a narrow intrinsic linewidth also as an important prerequisite for active stabilization.

Bulk dual-frequency fiber lasers and solid state lasers have been demonstrated \cite{ChenIEEETransMicrowTheoryTech2006, AlouiniIEEEPhotonicsTechnolLett1998}, the latter yielding a full-width at half-maximum (FWHM) linewidth of the beat frequency below 10 kHz. Aiming on chip-scale sources, dual-wavelength rare-earth-doped waveguide lasers have been widely explored as well \cite{GrivasProgQuantumElectron2016}, but these lasers require optical pumping, which introduces additional complexity. It is therefore attractive to use semiconductor lasers, which are directly pumped by an electric current. Semiconductor dual-wavelength lasers have been studied in the form of DBR lasers with a single gain section and a partly shared cavity \cite{PozziIEEEPhotonicsTechnolLett2006} and Bragg lasers with separate gain sections and separate laser cavities combined with a Y-junction \cite{PriceIEEEPhotonicsTechnolLett2007}, the latter showing a high output power of 70 mW.

The main problem of all on-chip dual-frequency semiconductor lasers realized so far is that the intrinsic linewidth of the individual laser frequencies is broad, e.g., around 60 MHz in \cite{PozziIEEEPhotonicsTechnolLett2006}, essentially due to a short photon lifetime in combination with high gain-index coupling \cite{HenryIEEEJQuantumElectron1982}. A well-known solution is to increase the photon lifetime by increasing the cavity length using low-loss waveguides. In this direction, there have been studies on heterogeneous integrated III-V/silicon dual-frequency lasers \cite{ShaoOptLett2014, HulmeOptExpress2017}. In\cite{ShaoOptLett2014}, a single laser cavity is used to generate a beat frequency at 0.357 THz with a FWHM linewidth of 4.2 MHz. In \cite{HulmeOptExpress2017}, two separate laser cavities and a photodetector are integrated on the same chip and used to generate beat frequencies between 1 and 112 GHz. Intrinsic linewidths of the individual lasers of 148 kHz are reported. The narrowest FWHM microwave linewidths so far were achieved using monolithic InP extended cavity lasers based on arrayed waveguide gratings (AWGs) and reached 130 kHz \cite{CarpinteroOptLett2012}, 100 kHz \cite{CarpinteroJLightTechnol2014}, and 56 kHz \cite{Guzman2014}. However, in all these approaches the photon lifetime, and thus the linewidth, is ultimately limited by the use of semiconductor waveguides, which have relatively high linear and nonlinear loss \cite{HeckLaserPhotonicsRev2014, XiangOptExpress2020}.

\begin{figure}[bp]
    \centering
    {\includegraphics[width=0.8\linewidth]{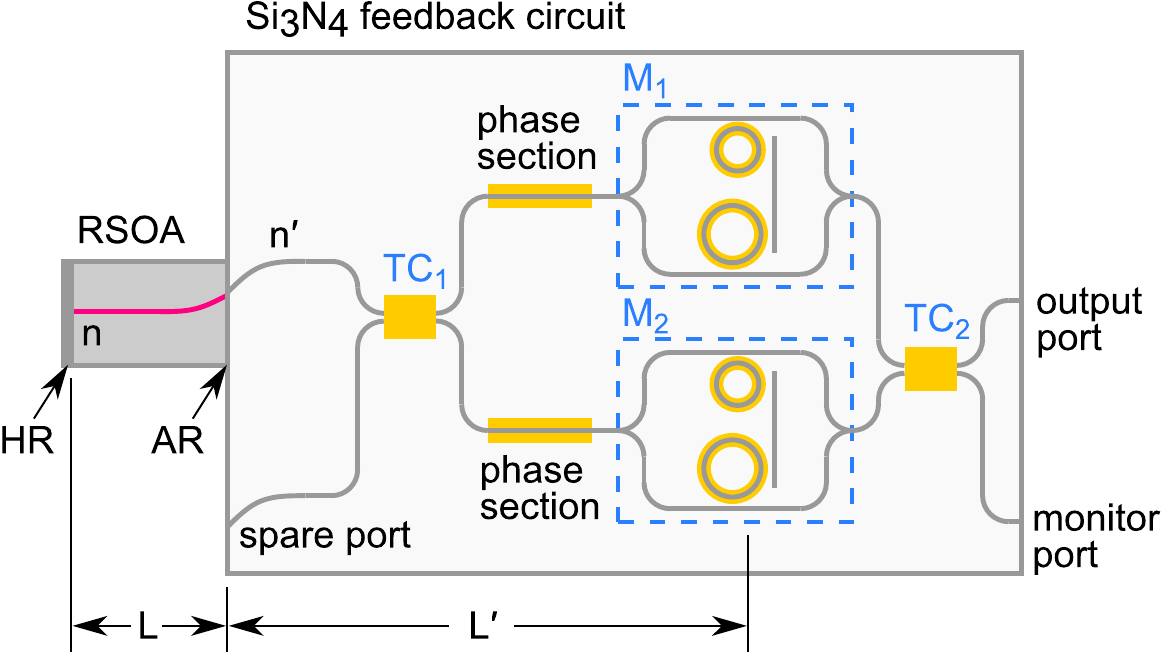}}
    \caption{Schematic diagram of the dual-frequency hybrid laser (not to scale). The reflective semiconductor optical amplifier (RSOA) is equipped on one facet with a high-reflection (HR) coating, and on the other facet with an anti-reflection (AR) coating. The silicon nitride (Si$_3$N$_4$) feedback chip contains two tunable couplers (labeled TC\textsubscript{1} and TC\textsubscript{2}), two phase sections, and two Vernier mirrors (labeled M\textsubscript{1} and M\textsubscript{2}) providing frequency-selective feedback. Each Vernier mirror consists of two microring resonators, which have slightly different ring radii $R_1$ = 125 $\mu$m and $R_2$ = 127.5 $\mu$m. The tunable couplers, phase sections, and ring resonators can be tuned using resistive electric heaters, indicated in yellow.}
    \label{fig:circuit}
\end{figure}
Here, we take an important step beyond such limits by extending the cavity photon lifetime using a low-loss dielectric feedback circuit via hybrid integration. Specifically, we use an indium phosphide (InP) amplifier hybrid integrated with a long and low-loss silicon nitride (Si$_3$N$_4$) feedback circuit (Fig. \ref{fig:circuit}), which extends the optical roundtrip length of the solitary amplifier from about 0.5 cm to 30 cm, i.e., a large factor of about 60. To enable dual-frequency lasing, we introduce a tunable coupler which directs the light towards two frequency-selective loop mirrors that can be controlled independently. Importantly, these mirrors are based on tunable microring resonators in Vernier configuration, as this allows also for wide wavelength coverage in single-frequency operation and enables continuous (mode-hop free) tunability over a wide range \cite{VanReesOptExpress2020}. We use the dual-frequency laser to generate narrow linewidth microwave radiation at a frequency of 11 GHz, and observe a narrow intrinsic linewidth of around 2 kHz.

We note that generating continuous-wave narrow intrinsic linewidth radiation is promising with quantum cascade lasers (QCLs) as well, but their room temperature operation as desired is restricted so far to much higher frequencies in the far-infrared region above 10 THz and the mid-IR region \cite{VitielloOptExpress2015}. A most convincing solution to access also the THz region at room temperature is found in intra-cavity difference frequency generation \cite{ConsolinoSciAdv2017}.

Closest to our study is the recent demonstration of two lasers using two separate semiconductor optical amplifiers of different materials (InP and GaAs) that are hybrid integrated with seperate feedback circuits on the same dielectric (Si$_3$N$_4$) chip \cite{ZhuOptExpress2019}. The aim of that study was to generate light in two largely different wavelength ranges (1.5 and 1 $\mu$m), with the future goal of generating mid-IR radiation via difference frequency generation. The individual lasers had relatively low intrinsic linewidths of 18 and 70 kHz. 

In contrast to that study and other previous work \cite{HulmeOptExpress2017}, we generate the two frequencies from a single gain medium, because this can enable additional linewidth reduction \cite{ChenIEEETransMicrowTheoryTech2006, PozziIEEEPhotonicsTechnolLett2006}. The reason is that the frequency fluctuation of the two modes may experience some synchronization if they are initiated by a common cause. However, this effect is difficult to investigate for semiconductor lasers. First, semiconductor lasers typically suffer from strong gain competition due to homogeneous broadening of the gain medium, which makes it difficult to establish simultaneous oscillation at two frequencies. Single-gain dual-frequency operation therefore requires an additional control parameter that balances the relative feedback. Second, noise correlation effects at fast noise frequencies are more complex for semiconductors. On the one hand, one expects that the two laser modes undergo independent phase excursions due to independent spontaneous emission noise. On the other hand, in semiconductors specifically, most of the intrinsic linewidth is due to index fluctuations associated with inversion fluctuations \cite{HenryIEEEJQuantumElectron1982}, which should still be seen by both oscillating frequencies. In hybrid and heterogeneously integrated lasers with spectrally narrow feedback, additional effects influence the intrinsic linewidth, specifically an all-optical feedback loop based on the coupling between feedback strength and index, which is usually expressed as the so-called B-factor \cite{VahalaApplPhysLett1984, KomljenovicApplSci2017, TranIEEEJSelTopQuantumElectron2020}.

Here, we present a demonstration of dual-frequency oscillation in the presence of gain competition, in which we achieve a record-low intrinsic linewidth of the beat of these frequencies of about 2 kHz. To our best knowledge, this is the narrowest linewidth reported so far for microwaves generated using a free-running chip-integrated dual-frequency laser. This narrow intrinsic linewidth enables a first view on correlation of frequency noise that can be seen in part of the beat spectra. Thereby, this work demonstrates novel options for on-chip high-purity microwave and THz generation. In particular, these lasers are of promise for further linewidth reduction as can be seen from recent measurements at single frequencies \cite{TranIEEEJSelTopQuantumElectron2020, FanOptExpress2020}.

\section{Methods}
A schematic diagram of the dual-frequency laser is shown in Fig. \ref{fig:circuit}. The laser consists of a double-pass reflective semiconductor optical amplifier (RSOA) chip, which generates light at a wavelength around 1.5 $\mu$m, and a Si$_3$N$_4$ chip, which extends the cavity length and provides frequency-selective feedback for dual-frequency lasing. The RSOA chip \cite{DeFelipeIEEEPhotonicsTechnolLett2014}, fabricated by the Fraunhofer Heinrich Hertz Institute, contains an InP-based multi-quantum well active waveguide with a length of $L$ = 700 $\mu$m and a group index of $n$ = 3.6. One facet is equipped with a high-reflection (HR) coating (reflectance 90\%) and forms one of the cavity mirrors. To minimize back reflections into the guided mode of the RSOA, the other facet (at the Si$_3$N$_4$-InP interface) is equipped with an anti-reflection (AR) coating. In addition, near the interface the amplifier waveguide is angled with respect to the facet normal by 9 degrees.

The Si$_3$N$_4$ feedback circuit is, in contrast to previously reported hybrid lasers \cite{FanOptExpress2020, MakOptExpress2019, VanReesOptExpress2020}, fabricated with an asymmetric double stripe (ADS) cross section. Compared to symmetric double stripe (SDS) waveguides, these require less fabrication steps and offer similar or lower propagation loss (< 0.1 dB/cm) \cite{RoeloffzenIEEEJSelTopQuantumElectron2018}. The ADS Si$_3$N$_4$ waveguides have a group index ($n'$) of 1.77. For optimum coupling efficiency, near the interface, the Si$_3$N$_4$ waveguide is tilted by about 20 degrees with respect to the facet normal. Two-dimensional tapering \cite{RoeloffzenIEEEJSelTopQuantumElectron2018} was used for optimized matching to the mode field of the RSOA.

To enable lasing at two frequencies, which we refer to as $f_1$ and $f_2$, the circuit contains two Vernier loop mirrors (M\textsubscript{1} and M\textsubscript{2}). These loop mirrors have a similar design as in \cite{MakOptExpress2019, VanReesOptExpress2020}, with two microring resonators (MRRs) which have slightly different radii ($R_1$ = 125 $\mu$m, $R_2$ = 127.5 $\mu$m). The power coupling to the bus waveguides was designed to be 1\%, to realize significant cavity length enhancement at resonance (enhancement factor $F$ = 99), due to multiple passes through the MRRs \cite{LiuApplPhysLett2001}. In total, the optical roundtrip length of the hybrid laser cavity is about 30 cm, i.e., a factor of about 60 longer than the roundtrip length of the solitary amplifier ($2 n L \approx$ 0.5 cm).

To enable dual-frequency oscillation despite the presence of gain competition, we have incorporated a Mach-Zehnder interferometer (MZI) based tunable coupler (labeled TC\textsubscript{1}) to balance the feedback levels from M\textsubscript{1} and M\textsubscript{2}. We note that the tunability of the feedback balance during laser operation is an important control parameter that also allows to investigate a possible influence of shared index fluctuations in the RSOA on the linewidth of the microwave frequency. To combine the laser output behind both Vernier mirrors into a single output port, the signals are superimposed using another tunable coupler (TC\textsubscript{2}).

All MRRs are equipped with thermo-electric heaters to enable tuning of the Vernier peaks, which makes it possible to set the laser to different light frequencies, $f_1$ and $f_2$. To set $f_1$, for example, to a desired value close to $f_2$, we simultaneously tune the two MRRs belonging to M\textsubscript{1} at the correct ratio, given by the ratio of MRR lengths. This ratio was experimentally found to be 1.03 for both M\textsubscript{1} and M\textsubscript{2}, which matches well with the expected value based on the design parameters ($R_2/R_1$ = 1.02). We recall that continuous mode-hop free tuning of a single-frequency version of the laser was demonstrated to span a range of 28 GHz \cite{VanReesOptExpress2020}. This suggests that with a dual-frequency laser mode-hop free tuning of the beat frequency over a range of 56 GHz would be possible. 

The RSOA chip and Si$_3$N$_4$ chip are hybridly integrated, and permanently fixed to a mount that is kept at a constant temperature (25 \textdegree{}C) using a Peltier cooler. The RSOA current and heater voltages are provided using USB-controlled printed circuit boards. To characterize the dual-frequency laser, the output is guided through an optical isolator to a 50:50 fiber splitter. 50 percent of the output power is guided to an optical spectrum analyzer (OSA; Ando AQ6317, spectral resolution 1.25 GHz). To generate and inspect beat notes directly in the microwave domain, the remaining light is guided to a fast photodector (Discovery Semiconductors DSC30S, 20 GHz bandwidth) connected to an electrical spectrum analyzer (ESA; Agilent E4405B).

\section{Results}
Without applying a voltage to TC\textsubscript{1}, in spite of a symmetric design, the feedback levels of the two Vernier mirrors (M\textsubscript{1} and M\textsubscript{2}) are likely to be imbalanced, as can be caused by inherent fabrication error in the MZI arm path lengths. Indeed, without a voltage applied to TC\textsubscript{1}, the hybrid laser generally shows single-frequency oscillation due to spectral condensation. The laser output power is about 1.5 mW at a pump current of 100 mA. The highest output power we observed was 5.5 mW using a pump current of 200.6 mA. With synchronized heating of the MRRs as described above, each wavelength can be independently controlled with a spectral coverage of 80 nm, which is close to the gain bandwidth of the diode.
\begin{figure}[t!]
    \centering
    {\includegraphics[width=\linewidth]{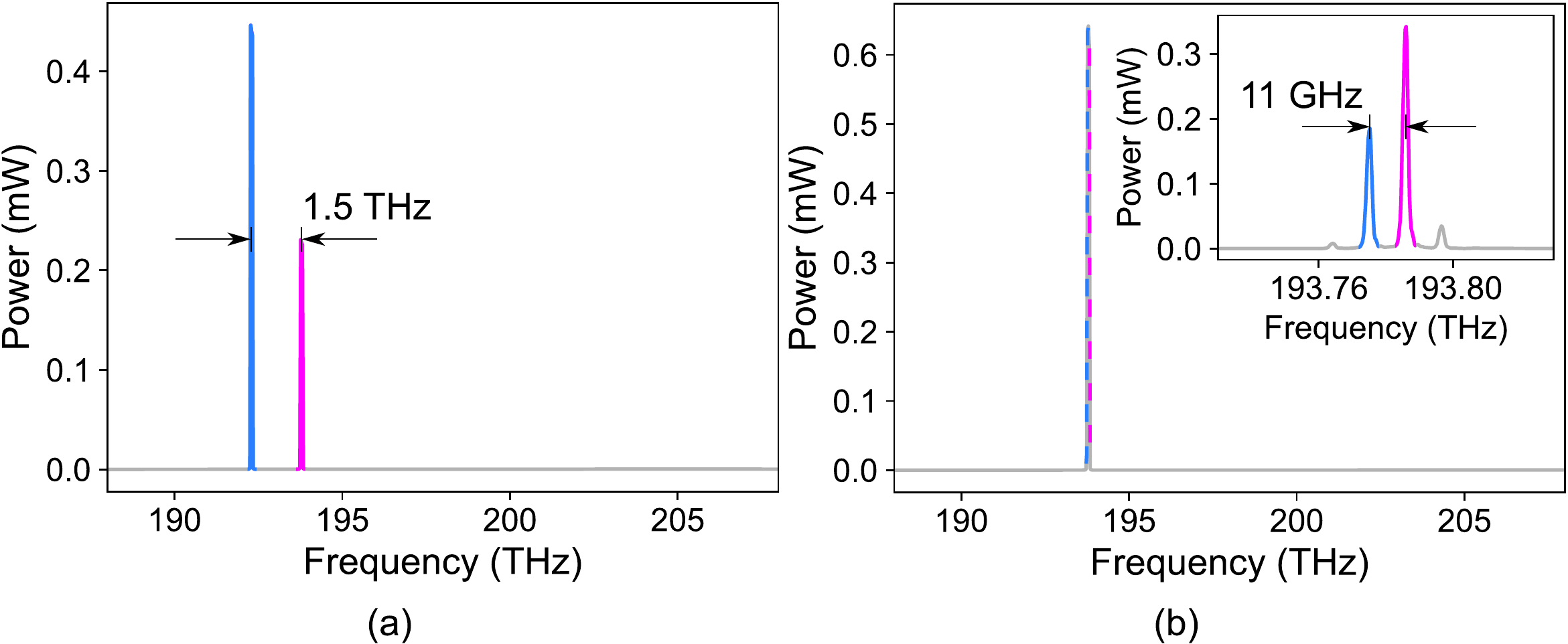}}
    \caption{Measured dual-frequency optical spectra with the two laser lines (highlighted in blue and pink) tuned to a frequency difference of (a) 1.5 THz (12 nm), and (b) 11 GHz (0.09 nm). Both (a) and (b) provide a wider overview at lower resolution (125 GHz). The inset of (b) shows a narrower frequency window with higher resolution (1.25 GHz), where the two laser lines can be seen, as well as two higher order sidebands.}
    \label{fig:optical_spectra}
\end{figure}

Dual-frequency operation is achieved by carefully balancing the feedback level using TC\textsubscript{1}. Figure \ref{fig:optical_spectra}(a) shows a typical example of the dual-frequency laser spectrum (OSA resolution 125 GHz), where the two frequencies ($f_1$ and $f_2$) are tuned to a frequency difference of 1.5 THz (12 nm). For microwave generation which can be easily measured using the ESA, the laser lines are tuned close to each other, here to a distance of 11 GHz (0.09 nm). This is shown in Fig. \ref{fig:optical_spectra}(b). The figure displays a wider overview of the optical spectrum at low resolution (125 GHz) and a narrower window at higher resolution (1.25 GHz) in the inset, where the two laser lines can be distinguished, as well as two higher order sidebands.

\begin{figure}[t!]
    \centering
    {\includegraphics[width=\linewidth]{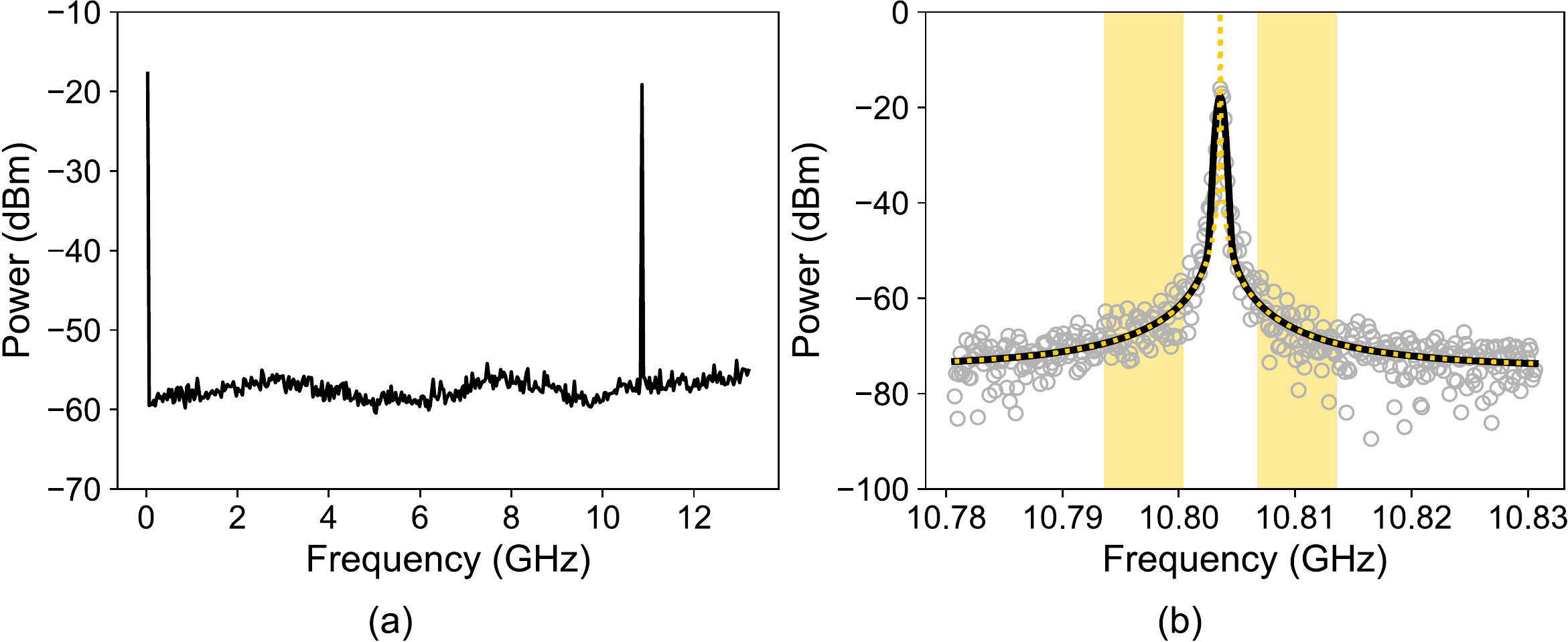}}
    \caption{(a) Beat frequency near 11 GHz, measured using a fast photodetector and an electrical spectrum analyzer (ESA) set to a resolution and video bandwidth of 3 MHz each. (b) Close-up example of the measured beat frequency (open circles), recorded with a resolution and video bandwidth of 300 kHz each. The solid line represents a Voigt fit with a Gaussian (technical noise) component $w_G$ = 0.42 MHz and a Lorentzian component $w_L$ = 5.4 kHz (FWHM). Here, $w_G$ was a free fit parameter, while $w_L$ was kept fixed and was determined by fitting a Lorentzian (yellow dashed line). As in \cite{FanOptExpress2020}, the fit is based only on data in the center of the wings (shaded area), which reduces biasing by non-Lorentzian components from the Gaussian line center and flat noise floor.}
    \label{fig:beat_note_spectra}
\end{figure} 
\begin{figure}[ht!]
    \centering
    {\includegraphics[width=\linewidth]{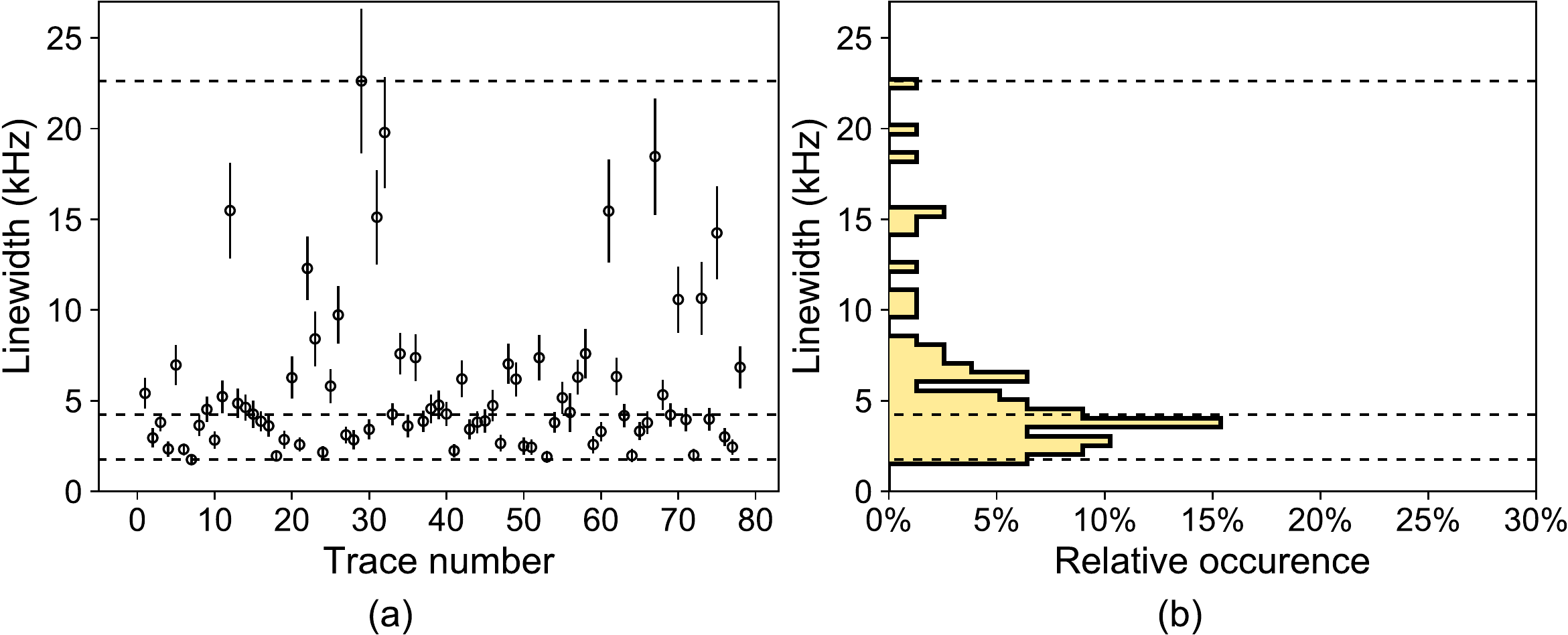}}
    \caption{(a) Intrinsic linewidths of the generated 11-GHz microwave signal obtained with longer term recording of the free-running dual-frequency laser output. The vertical error bars represent the 95\% confidence intervals calculated from the standard error of the fit parameter. (b) Distribution of the measured linewidth values. The horizontal dashed lines indicate the minimum, median, and maximum linewidth values.}
    \label{fig:beat_note_linewidth}
\end{figure}
To demonstrate narrow-linewidth microwave generation, we guided the dual-frequency output to a fast photodetector and recorded the electrical spectrum, which is shown in Fig. \ref{fig:beat_note_spectra}(a). The spectrum displays a sharp peak near 11 GHz, which matches the frequency difference in the optical spectrum (Fig. \ref{fig:optical_spectra}(b)). To inspect the intrinsic linewidth, we recorded multiple microwave spectra using a narrower frequency window (50 MHz). The center frequency and spectral shape showed some variation over time due to remaining technical noise, e.g., from acoustic fluctuations or noise in the pump current. To avoid developing an active stabilization for removal of slower variations, we characterized the non-stabilized laser (free-running) with repeated recordings of sufficiently short aquisition time (4 ms).

Figure \ref{fig:beat_note_spectra}(b) shows a representative example of the generated microwave spectra. The spectrum follows a Voigt profile (solid line in Fig. \ref{fig:beat_note_spectra}(b)), with a Gaussian-shaped center indicating technical noise, and Lorentzian-shaped wings due to the intrinsic optical linewidths of the laser. To determine the intrinsic linewidth of the microwave spectra, we applied a Lorentzian least-square fit (dashed yellow line) to the center of the wings, which reduces biasing by the Gaussian-shaped line center and the flat noise floor \cite{FanOptExpress2020}. The Lorentzian width of the spectra, which sets the intrinsic linewidth, showed some variation over time. This is likely caused by a residual drift of the light frequencies with regard to the center frequencies of the Vernier filters. Such drift can reduce the number of roundtrips in the MRRs, thereby shortening also the laser's effective cavity length. To provide information not only on the narrowest linewidth that can be achieved, we performed longer-term linewidth recording, i.e., we performed 80 recordings of the beat note in sequence, without realignment, with an interval of a few seconds between each acquisition.

Figure \ref{fig:beat_note_linewidth}(a) shows the intrinsic linewidth values (FWHM) that we obtained for the 80 spectra. The error bars represent the 95\% confidence intervals calculated from the standard error in the fitted linewidth. Figure \ref{fig:beat_note_linewidth}(b) shows the same data as a histogram. The horizontal dashed lines in Figs. \ref{fig:beat_note_linewidth}(a) and (b) indicate the minimum, median, and maximum linewidths. To provide an indication of the typical linewidth value, we chose to present the median value rather than the average, because the average (5.7 kHz) is slightly skewed by a small number of larger linewidth values. The broadest linewidth that we observed is 23 $\pm$ 4 kHz. The median linewidth value is 4.2 kHz. The lowest observable linewidth was 1.8 $\pm$ 0.3 kHz, which is, to our best knowledge, the narrowest intrinsic linewidth obtained so far for microwave generation using free-running chip-integrated dual-frequency lasers.

To pursue also the question whether generating the two frequencies from a single gain section allows for microwave linewidths narrower than the convolution of the individual laser frequencies, we measured also the intrinsic linewidth of the individual laser frequencies. As we are investigating intrinsic linewidth components of Lorentzian shape, this convolution can be expressed by adding the optical linewidths.

\begin{figure}[b!]
    \centering
    {\includegraphics[width=\linewidth]{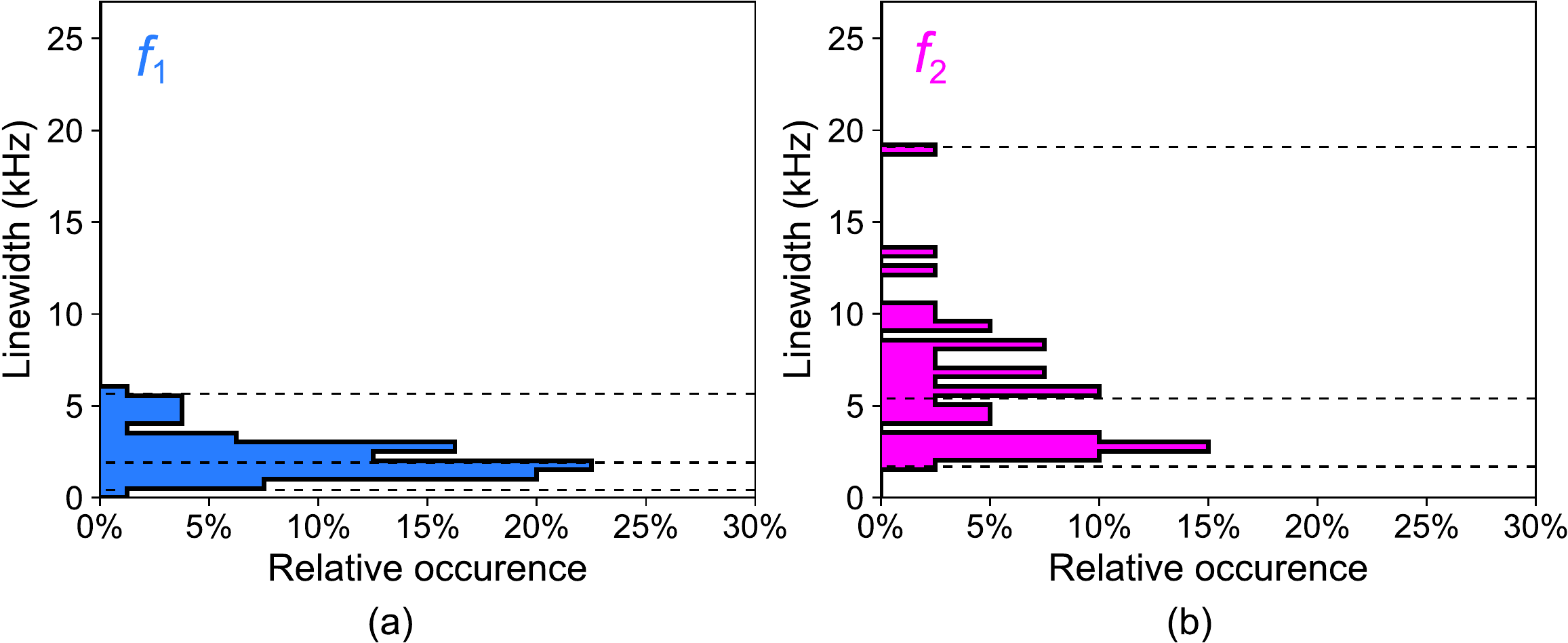}}
    \caption{Distribution of the measured intrinsic linewidths for (a) the laser line at $f_1$, and (b) the laser line at $f_2$, each after collapse to single-frequency oscillation. The linewidths were obtained from self-heterodyne beat spectra recorded using an electrical spectrum analyzer (ESA). The minimum, median, and maximum linewidth values are indicated by the horizontal dashed lines.}
    \label{fig:individual_linewidth_histograms}
\end{figure}

To enable measuring the individual optical linewidths, we temporarily imbalanced TC\textsubscript{1} until the spectrum collapsed to a single frequency, first at $f_1$ and then at $f_2$, and adjusted the phase section for narrowest linewidth. We measured the linewidth via delayed self-heterodyne detection \cite{OkoshiElectronLett1980} using the experimental setup described in \cite{FanIEEEPhotonicsJ2016}, again recording multiple electrical spectra and fitting a Lorentzian to the center of the wings for each spectrum. Using a 50 MHz window, corresponding to a resolution and video bandwidth of 300 kHz each, we found the linewidth distributions shown in Figs. \ref{fig:individual_linewidth_histograms}(a) and (b). As indicated by the dashed lines, the measurements yielded two somewhat different median linewidth values for the two laser frequencies, namely 1.9 kHz (at $f_1$) and 5.4 kHz (at $f_2$). This may be addressed to the two Vernier feedback circuits providing a somewhat different extension of the optical length of the resonator, which enters the intrinsic linewidth approximately quadratically \cite{FanOptExpress2020}. The difference may have been caused by small deviations in fabrication, specifically in the exact strength of the couplers at the MRRs. The same trend of a narrower linewidth with the $f_1$-feedback circuit was found in all measurements. The minimum individual linewidths that were observable are as low as 0.4 $\pm$ 0.1 kHz (at $f_1$) and 1.7 $\pm$ 0.2 kHz (at $f_2$). 

To interpret the optical linewidth data in context of the measured microwave linewidths, we recall that if correlations in the frequency fluctuations of the two laser lines are absent, the microwave linewidth should be given by the sum of the two individual Lorentzian linewidths. Looking at the median individual linewidths, one would then expect a microwave linewidth of 1.9 + 5.4 = 7.3 kHz. However, the measured median microwave linewidth (4.2 kHz) is clearly smaller than 7.3 kHz. This observation suggests that the two laser lines show correlated frequency fluctuations. However, when comparing the minimum linewidth values, the microwave linewidth (1.8 $\pm$ 0.3 kHz) is, within the experimental error, the same as the added individual linewidths (2.1 $\pm$ 0.3 kHz). A possible explanation is that the minimum linewidth values occured for near-optimum alignment of the light frequencies with respect to the Vernier center frequencies, i.e., slightly detuned, which leads to compensation of inversion induced index and frequency fluctuations, expressed via the B-factor \cite{VahalaApplPhysLett1984, KomljenovicApplSci2017, TranIEEEJSelTopQuantumElectron2020}. Although further research is certainly required, it appears that we have found traces of a beneficial correlation in the fast frequency fluctuations of a dual-frequency diode laser caused by the use of a single gain element.

\section{Summary and conclusion}
In summary, we have presented a hybrid integrated dual-frequency laser for microwave and THz generation. With the example of microwave generation at 11 GHz, we observe extremely low intrinsic linewidths with a median value of 4.2 kHz in measurement series and 1.8 $\pm$ 0.3 kHz (smallest linewidth). These values are, to our knowledge, the narrowest linewidths reported so far for microwaves generated by a free-running chip-integrated dual-frequency laser. This is achieved by hybrid integrating an InP amplifier with a low-loss dielectric (Si$_3$N$_4$) feedback ciruit, which extends the optical roundtrip length by a large factor of about 60 with respect to the solitary diode, i.e., from 0.5 cm to about 30 cm on a chip, and provides feedback for dual-frequency lasing via two independently tunable, frequency-selective Vernier mirrors.

Our approach can provide tunable microwave and THz generation at a wide range of frequencies, from low frequencies in the RF or microwave range up to 10 THz, because the InP-Si$_3$N$_4$ lasers that were used here have a large spectral coverage, essentially covering the entire gain bandwidth \cite{VanReesOptExpress2020}. The narrow linewidth we have demonstrated should also be found back in the THz range when applying the appropriate linewidth analysis \cite{DucournauOptLett2011}. In addition, mode-hop free tunability in the microwave and THz regime could be implemented straightforwardly via synchronous tuning of the microring resonators and phase sections as was demonstrated recently with a similar, single-frequency InP-Si$_3$N$_4$ hybrid integrated laser \cite{VanReesOptExpress2020}. The Si$_3$N$_4$ platform used in our approach enables attractive options for microwave photonics, as it enables co-integration with advanced components such as modulators and filters \cite{MarpaungNatPhotonics2019}.

Future studies should focus on further clarification and possibly exploitation of correlations in the frequency noise of dual-frequency, shared amplifier hybrid integrated diode lasers for further linewidth narrowing in the microwave and THz range. We note that even if such noise correlation turns out to be limited, there lies significant potential for further linewidth narrowing by using longer feedback circuits \cite{FanOptExpress2020}, potentially resulting in microwave and THz linewidths below the 100 Hz level.

\section{Funding}
Netherlands Organization for Scientific Research (NWO) (13537); H2020 LEIT Information and Communication Technologies (3PEAT, 780502); Rijksdienst voor Ondernemend Nederland (RVO) (IOP Photonic Devices program).

\section{Acknowledgments}
We thank C. A. A. Franken, H. M. J. Bastiaens, and D. Marpaung for support and suggestions.

\bibliography{references}

\end{document}